\documentclass[12pt]{article}

\usepackage{mathrsfs}
\usepackage{verbatim}
\usepackage{amsmath}
\usepackage{hyperref}
\usepackage{amssymb}
\usepackage{stmaryrd}
\usepackage{tensor}
\usepackage{graphicx}
\usepackage{amsthm}
\usepackage{braket}
\usepackage{accents}
\usepackage{bm}
\usepackage{tikz-cd}
\newcommand{\mbb}{\mathbb}
\newcommand{\ts}{\ \,}

\newcommand{\prol}[1]{j_s#1}
\newcommand{\prolc}[1]{j_c#1}
\newcommand{\pord}{W}
\newcommand{\spoint}{z}

\newtheorem{mydef}{Definition}

\usepackage{authblk}
\makeatletter
\renewcommand*\env@matrix[1][\arraystretch]{%
  \edef\arraystretch{#1}%
  \hskip -\arraycolsep
  \let\@ifnextchar\new@ifnextchar
  \array{*\c@MaxMatrixCols c}}
\makeatother

\newenvironment{sciabstract}{%
\begin{quote} \bf}
{\end{quote}}

\title{Generalized conformal maps as classical symmetries of Yang-Mills fields}

\author{Edward B. Baker III\thanks{edwardbaker86@gmail.com}}

\date{\today}

\begin{document} 

\maketitle 

\begin{sciabstract}

We show that a class of previously defined maps, called causal and self-dual morphisms, form classical symmetries of Yang-Mills fields in four complex dimensions.  These maps generalize conformal transformations, and admit a nonlocal pullback connection that preserves the equations of the theory.  First it is shown that self-dual morphisms form symmetries of the anti-self-dual Yang-Mills equations under this pullback.  Then a supersymmetric generalization of causal morphisms is defined which preserves solutions of the field equations for N=3 supersymmetric Yang-Mills theory.  As a special case, this implies that a modified definition of causal morphisms form symmetries for the ordinary Yang-Mills field equations.
\end{sciabstract}

\section{Introduction}

Hidden symmetries have played an important role in the study of Yang-Mills (YM) theory. As an example, the anti-self dual Yang-Mills (ASDYM) equations have an infinite class of hidden symmetries which bear some resemblance to the infinite-dimensional conformal group in two dimensions \cite{DOLAN1982387}\cite{CHAU1983391}\cite{POPOV_1999}.  In addition, an extended conformal symmetry called dual superconformal invariance has been uncovered in the study of N=4 supersymmetric Yang-Mills (SYM) theory \cite{Mason_2009}\cite{Arkani_Hamed_2010}, leading to an infinite dimensional Yangian symmetry \cite{Drummond_2009}.  
 This and other advances have led to powerful tools for the study of N=4 SYM. 

Many of these results can be understood best with the use of twistor and ambitwistor methods, which have been used extensively in the study of Yang-Mills fields.  For example, the Penrose-Ward correspondence reformulates the ASDYM equations in Twistor space, which leads to the ADHM construction of instantons \cite{1977PhLA...61...81W}\cite{ATIYAH1978185}\cite{atiyah1979geometry}.  This construction was generalized to a geometric formulation of the Yang-Mills field equations in ambitwistor space, with a natural interpretation in superspace \cite{WITTEN1978394}\cite{ISENBERG1978462}\cite{Harnad1985Constraint}\cite{harnad1989supersymmetric}.  More recently, twistor and ambitwistor methods have been used in string theory to understand Yang-Mills scattering amplitudes and their properties \cite{Witten_2004}\cite{atiyah2017twistor}.

In this paper we investigate a previously defined class of generalized maps \cite{https://doi.org/10.48550/arxiv.2203.07952} in the context of Yang-Mills theory. These maps are motivated by  twistor and ambitwistor theory, and are called self-dual and causal morphisms, respectively.  Under certain assumptions, one can define a non-local pullback connection under these transformations that preserves integrability on certain subspaces.  In the case of self-dual morphisms, the maps preserve integrability on self-dual planes which imply that they are symmetries of the ASDYM equations.  A supersymmetric generalization of causal morphisms is then developed which preserves integrability on super null lines, implying that these maps are symmetries of the N=3 SYM field equations.  This fact is used to show that a modified version of causal morphisms are symmetries of the YM field equations as a special case.    It is likely that some of these symmetries are related to known hidden symmetries for the different cases, but characterizing these relationships will be left as a topic of future investigation.

\section{Self-dual morphisms as symmetries of ASDYM}

Self-dual morphisms were introduced in a previous paper, where they were defined using maps on null surfaces \cite{https://doi.org/10.48550/arxiv.2203.07952}.  Here we provide a self contained summary of these results, using different but equivalent definitions.   To begin, define the twistor correspondence space $\mathcal{F}=\mbb{C}^4\times \mbb{C}\mbb{P}^1$ with the usual double fibration \cite{dunajski2010solitons}\cite{ward1991twistor}
\begin{equation}\label{eq:doublefib}
\mbb{C}^4\xleftarrow{\pi_1} \mathcal{F} \xrightarrow{\pi_2} \mathcal{P}\mathcal{T},
\end{equation}
where $\mathcal{P}\mathcal{T}=\mbb{C}\mbb{P}^3$ is the projective twistor space of $\mbb{C}^4$.  Now define a self-dual embedding as a totally null holomorphic embedding $\chi:\mbb{C}^2\rightarrow \mbb{C}^4$ such that vectors at a point $t\in \mbb{C}^2$ are mapped under $\chi_\star$ to vectors of the form $v^{\alpha\dot{\alpha}}=\lambda^\alpha\tilde{\lambda}^{\dot{\alpha}}$ for $\tilde{\lambda}^{\dot{\alpha}}$ fixed.  We will call the image of such a map a self-dual surface.  If $\tilde{\lambda}$ is independent of $t$ then this surface maps to a self-dual plane (or $\alpha$-plane) $Z$. Furthermore, for any point on a self-dual embedding there is a tangent $\alpha$-plane passing through $\chi(t)$ that is characterized by $\tilde{\lambda}(t)$.  For brevity, we say that $\chi$ is tangent to $\tilde{\lambda}$ at $t$.  Now define the self-dual prolongation $\prol{\chi}:\mbb{C}^2\rightarrow \mathcal{F}$ by $\prol{\chi}=(\chi,\tilde{\lambda})$, where dependence on $t$ is suppressed.  The prolongation satisfies a contact condition, that $\chi$ is tangent to $\tilde{\lambda}$ for all $t\in \mbb{C}^2$.  Conversely, given a surface $\psi:\mbb{C}^2\rightarrow \mathcal{F}$, we say that it satisfies the contact condition if $\psi=\prol{\chi}$ for some self-dual embedding $\chi$.  A map $f:\mathcal{F}\rightarrow \mathcal{F}$ is said to preserve the contact condition if $f\circ \prol{\chi}$ satisfies the contact condition for any $\chi$.\footnote{These constructions are all assumed to be local and defined in some neighborhood, but are written globally for ease of notation.}  We  then define
\begin{mydef}
A self-dual morphism is a holomorphic map $f:\mathcal{F}\rightarrow \mathcal{F}$ which preserves the contact condition.
\end{mydef}
  Defined in this way, a self-dual morphism naturally induces maps on self-dual embeddings and self-dual planes
  \begin{mydef}
  Given a self-dual morphism $f$ and a self-dual embedding $\chi$, define the contraction map $f\lrcorner \chi := \pi_1\circ f\circ \prol{\chi}$.  Furthermore, for a self-dual plane $Z$ tangent to $\tilde{\lambda}$, define $f\lrcorner Z:Z\rightarrow \mbb{C}^4$ by $f\lrcorner Z(x)=\pi_1\circ f(x,\tilde{\lambda})$ where $x\in Z$. 
 \end{mydef}
This map on surfaces was the starting point for the definitions in the previous paper, and the two definitions are equivalent.


Now consider a $GL(n,\mbb{C})$ connection with vector potential $A$ satisfying the ASDYM equations on $M_\mbb{C}=\mbb{C}^4$.  Given a self-dual morphism $f$, there is a natural definition for a pullback connection $f^*A$.  To see this, first restrict to a self-dual plane $Z$, which can be parameterized linearly by coordinates on $\mbb{C}^2$.  The contraction map $f\lrcorner Z$ then gives a self-dual embedding,  and the pullback connection $(f\lrcorner Z)^*A$ is integrable on $Z$ because the curvature of $A$ vanishes on the self-dual planes tangent to $f\lrcorner Z$ as a consequence of ASDYM.  By varying $Z$ this allows us to define the bundle of parallel sections on the twistor space, and to use the Penrose-Ward procedure to define a connection on the pullback bundle, which defines the pullback connection $f^* A$ and gives a solution of ASDYM.  This requires that the bundle of parallel sections is trivial for points $x\in\mbb{C}^4$, which will be shown with an explicit construction of $f^*A$. 

For the construction, consider two points $x_1,x_2\in Z$ and their images  $y_i=f\lrcorner Z(x_i)$.  Define a Wilson line for the pullback connection by
\begin{equation}
\pord^*_Z (x_1,x_2)=\pord_{f\lrcorner Z}(y_1,y_2)=\mathcal{P}\exp\left(\int_{\gamma}A_\mu d x^\mu\right).
\end{equation}
Here the path of integration is any path $\gamma$ confined to the image of $f\lrcorner Z$ starting at $y_1$ and ending at $y_2$.  The Wilson line is independent of path due to the integrability of the connection on self-dual surfaces.  We can then define the patching matrix used in the Penrose-Ward correspondence by 
\begin{equation}
G=\pord^*_Z(q,p)=\tilde{H}H^{-1}
\end{equation}
where
\begin{equation}\label{eq:hdef}
H=\pord^*_Z(p,x), \ \ \tilde{H}=\pord^*_Z(q,x). 
\end{equation}
Here $p$ and $q$ are the points of intersection between $Z$ and self-dual planes $P$ and $Q$ with twistor coordinates $\hat{P}=(0,0,1,0)$ and $\hat{Q}=(0,0,0,1)$, as defined in the usual patching construction.  The patching matrix descends to the twistor space, and the splitting formula guarantees that the bundle is trivial, so the bundle satisfies the conditions of the Penrose-Ward correspondence.  We therefore can recover a self-dual pullback connection $f^*A$. This all assumes that the integrals are non-singular, which depends on the details of the maps, connections and domains under consideration.

To gain intuition for this construction, it is useful to derive an explicit formula for $f^*A$.  To this end, restrict to a self-dual plane $Z$ and use the formula  $\tilde{\lambda}^{\dot{\alpha}}f^*A_{\alpha\dot{\alpha}}=H^{-1}\tilde{\lambda}^{\dot{\alpha}}\partial_{\alpha\dot{\alpha}}H$ from the Penrose-Ward correspondence, in addition to equation \eqref{eq:hdef}, to find 
\begin{equation}\label{eq:apullback}
v \cdot f^*A\lvert_{x}=(f\lrcorner Z)_*v \cdot A\lvert_{f\lrcorner Z(x)}, \ \ \  x\in Z, \ v\in T_xZ.
\end{equation}
By varying the self-dual plane through a fixed point $x$ this gives the value for null vectors $v\in T_x\mbb{C}^{4}$, and the value on other vectors can be recovered from linearity of the connection.  Linearity and self-duality follow from the Penrose-Ward construction, but it is instructive to derive these results directly from equation \eqref{eq:apullback}, which can be expanded as
\begin{equation}
(f\lrcorner Z)_*v\cdot A\lvert_{f\lrcorner Z(x)}=v^{\alpha\dot{\alpha}}\Big(\frac{\partial y^{\beta\dot{\beta}}}{\partial x^{\alpha\dot{\alpha}}}A_{\beta\dot{\beta}}\Big)\Big\lvert_{f\lrcorner Z(x)}, \ x\in Z, \ v\in T_xZ,
\end{equation}
where $y^{\beta\dot{\beta}}=f\lrcorner Z(x)^{\beta\dot{\beta}}$. This equation is degree one in both $\lambda$ and $\tilde{\lambda}$ and globally holomorphic on $CP^1\times CP^1$, and so is linear by Liouville's theorem. Self-duality is equivalent to the connection being integrable on self-dual planes, which follows from using equation \eqref{eq:apullback} to show that $[v_1\cdot D^*,v_2\cdot D^\star]=0$, where $v_{1,2}\in T_x Z$ are linearly independent vectors tangent to a self-dual plane $Z$.  These arguments generalize well to the $N=3$ supersymmetric case.

\section{Super causal morphisms and N=3 SYM}

In this section we will define a super causal morphism, which is an extension of the previously defined causal morphisms to superspace.  The supersymmetric generalization is useful because of an interpretation of the N=3 SYM field equations as an integrability condition on supersymmetric null lines \cite{WITTEN1978394}\cite{Harnad1985Constraint}.  This interpretation allows for a generalization of the arguments of the previous section to the $N=3$ SYM field equations.  Furthermore, solutions of the usual YM field equations are special cases of the supersymmetric solutions, and this will be used to show that a modified version of causal morphisms are also symmetries of the ordinary YM field equations. 

The definition of super causal morphisms follows closely to the previous definitions. To begin, consider the superspace $\mbb{C}^{4\vert 4N}$ with coordinates $\spoint^A=(x^{\alpha\dot{\alpha}},\theta^{i\alpha},\tilde{\theta}_j^{ \dot{\alpha}})$ and supersymmetry generators $q_{i\alpha}=\frac{\partial}{\partial \theta^{i\alpha}}+i\tilde{\theta}^{\dot{\alpha}}_i\frac{\partial}{\partial x^{\alpha\dot{\alpha}}}$ and $\tilde{q}_{\dot{\alpha}}^i=\frac{\partial}{\partial \tilde{\theta}^{\dot{\alpha}}_i}+i\theta^{i\alpha}\frac{\partial}{\partial x^{\alpha\dot{\alpha}}}$.   The lightlike lines through $\spoint\in\mbb{C}^{4\vert 4N}$ and tangent to $(\lambda,\tilde{\lambda})$ are generated by fermionic translation operators $T_i=\lambda^\alpha q_{\alpha i}$ and $\tilde{T}^i=\tilde{\lambda}^{\dot{\alpha}}\tilde{q}_{\dot{\alpha}}^{i}$ which satisfy the algebra 
\begin{equation}\label{eq:symintegrability}
\{T_{i},T_{j}\}=\{\tilde{T}^{i},\tilde{T}^{j}\}=0, \ \ \{T_{i},\tilde{T}^{j}\}=2i\delta_i^j D,
\end{equation}
where $D=\lambda^\alpha\tilde{\lambda}^{\dot{\alpha}}\partial_{\alpha\dot{\alpha}}$.  Define the correspondence space $(\mathcal{F}^{6\vert 4N},\pi_1,\mbb{C}^{4\vert 4N})$ as the bundle over superspace whose fiber at a point $\spoint$ is the set of super null lines that intersect $\spoint$.  These fibers are isomorphic to $CP^1\times CP^1$, corresponding to the projective spinors $\lambda$ and $\tilde{\lambda}$ that generate bosonic translations along a given null line.  

The super null lines described above have one complex dimension and $2N$ fermionic dimensions, which can be parameterized by coordinates $\sigma=(s,\xi^i, \tilde{\xi}_j)\in \mbb{C}^{1\vert 2N}$.  
The supersymmetry generators on this line are $\partial_s$, $q_i=\partial_{\xi^i}+i\tilde{\xi}_i\partial_s$ and $\tilde{q}^i=\partial_{\tilde{\xi}_i}+i\xi^i\partial_s$.  More generally, we can also consider a super null curve, which is a morphism $\chi:\mbb{C}^{1\vert 2N} \rightarrow \mbb{C}^{4\vert 4N}$ whose pushforward $\chi_*$ takes the form \begin{equation}\label{eq:spush}
(\partial_s,q_i,
\tilde{q}^k)\rightarrow (D, M_i^{j}T_{j},\tilde{M}^k_{l}\tilde{T}^l),
\end{equation}
at every point $\sigma\in \mbb{C}^{1\vert 2N}$,
where the operators $(D, T_j, \tilde{T}^l)$ are defined for a super null line tangent to $\chi$ at $\chi(\sigma)$.  As with the self-dual case, we can then define the prolongation $\prolc{\chi}:\mbb{C}^{1\vert 2N}\rightarrow \mathcal{F}^{6\vert 4N}$ by $\prolc{\chi}=(\chi,\lambda, \tilde{\lambda})$,  and given a curve $\psi:\mbb{C}^{1\vert 2N}\rightarrow \mathcal{F}^{6\vert 4N}$ we say that it satisfies the contact condition if $\psi=\prolc{\chi}$ for some nonsingular null curve $\chi$.  As before, a map $f:\mathcal{F}^{6\vert 4N}\rightarrow \mathcal{F}^{6\vert 4N}$ preserves the contact condition if $f\circ \prolc{\chi}$ satisfies the contact condition for any $\chi$.  Furthermore, for a super null line $L$ tangent to $(\lambda,\tilde{\lambda})$, we also define $f\lrcorner L(\spoint)= \pi_1\circ f(\spoint,\lambda,\tilde{\lambda}) \  \forall \spoint\in L$ as before.
 
In the supersymmetric case there is an extra consideration necessary to ensure integrability of the pullback connection. To see this, consider a morphism $f:\mathcal{F}^{6\vert 4N}\rightarrow \mathcal{F}^{6\vert 4N}$ which preserves the contact condition.  Given a super null line $L$, we want to demand that the supersymmetry relations  \eqref{eq:symintegrability}
are preserved under $(f\lrcorner L)_*$.  By construction, this pushforward takes the form of equation \eqref{eq:spush}, where the coordinates on $L$ are also chosen to satisfy the supersymmetry relations.  To preserve these relations, we must demand
\begin{equation}
M^i_{j}\tilde{M}_k^{j}=\delta^i_k.
\end{equation}
This condition must be satisfied for every super null line $L$, which is the condition that must be satisfied for integrability.
 
 We can now define
\begin{mydef}
A super causal morphism is a holomorphic map $f:\mathcal{F}^{6\vert 4N}\rightarrow \mathcal{F}^{6\vert 4N}$ which preserves the contact condition, and preserves the supersymmetry relations \eqref{eq:symintegrability} under $(f\lrcorner L)_\star$ for tangent vectors to any super null line $L$.
\end{mydef}

Now consider an N=3 supersymmetric Yang-Mills field satisfying the field equations on $\mbb{C}^{4\vert 12}$.  This field can be defined by a superconnection characterized by a one form $\Phi$ with components $\Phi_A=(\omega_{i\alpha}, \tilde{\omega}^i_{\dot{\alpha}}, A_{\alpha\dot{\alpha}})$, which defines covariant derivative operators 
\begin{align}
Q_{i\alpha}=q_{i\alpha}+\omega_{i\alpha}, \ \tilde{Q}^i_{\dot{\alpha}}=q^i_{\dot{\alpha}}+\tilde{\omega}^i_{\dot{\alpha}}, \ D_{\alpha\dot{\alpha}}=\partial_{\alpha\dot{\alpha}}+A_{\alpha\dot{\alpha}}.
\end{align}
The field equations are equivalent to integrability on super null lines \cite{Harnad1985Constraint}, which for a given line $L$ tangent to $v^{\alpha\dot{\alpha}}=\lambda^{\alpha}\tilde{\lambda}^{\dot{\alpha}}$ are given by equations \eqref{eq:symintegrability} for the translation operators $T_{i}=\lambda^\alpha Q_{i\alpha}$,  $\tilde{T}^i=\tilde{\lambda}^{\dot{\alpha}}\tilde{Q}_{\dot{\alpha}}^{i}$ and $D=\lambda^{\alpha}\tilde{\lambda}^{\dot{\alpha}}D_{\alpha\dot{\alpha}}$.

Now, given a super causal morphism $f$, we can define the pullback connection $f^*\Phi$ similarly to the self-dual case.  To do so, consider a super null line $L$, and the super null curve $f\lrcorner L$ it generates.  The bundle of parallel sections on these super null lines then generates a pullback connection through the generalized Penrose-Ward construction.  To calculate this pullback connection, we can directly generalize equation \eqref{eq:apullback} to
\begin{equation}\label{eq:sympullback}
v \cdot f^*\Phi\lvert_{\spoint}=(f\lrcorner L)_*v \cdot \Phi\lvert_{f\lrcorner L(\spoint)}, \ \ \  \spoint\in L, \ v\in T_x L.
\end{equation}
As in the previous section, linearity of the pullback connection follows from a variant of Liouville's theorem.  Integrability on lines follows from writing \eqref{eq:symintegrability} for the pullback translation operators defined on $L$, and then using \eqref{eq:sympullback} and the assumption that $\Phi$ is integrable on lines.  This implies that super causal morphisms are symmetries of the $N=3$ SYM field equations.

\section{Reduction to Yang-Mills field equations}

The geometric interpretation of the YM field equations using field extensions can be naturally understood by viewing these equations as a special case of the N=3 SYM field equations for a Yang-Mills super multiplet with the scalar and spinor fields set to zero \cite{WITTEN1978394}\cite{Harnad1985Constraint}\cite{harnad1989supersymmetric}\cite{10.2307/2000661}. In a similar spirit, it is possible to use a modified definition of causal morphisms to generate an N=3 super causal morphism which preserves the property that the scalar and spinor fields equal zero, thus forming a symmetry of the YM field equations.  

A causal morphism can be defined as an $N=0$ super causal morphism, or a map $f:\mathcal{G}\rightarrow \mathcal{G}$ that preserves the contact condition, where $\mathcal{G}=\mathcal{F}^{6\vert 0}$ is the usual ambitwistor correspondence space.   Given such a function, we can construct an extended morphism $\hat{f}:\mathcal{F}^{6\vert 4N}\rightarrow\mathcal{F}^{6\vert 4N}$ given by
\begin{equation}\label{eq:extcaus}
\hat{f}(g,\theta^{i\alpha},\tilde{\theta}_j^{\dot{\alpha}})=(f(g),[V^{-1}]^\alpha_{\beta}\theta^{i\beta },[\tilde{V}^{-1}]^{\dot{\alpha}}_{ \dot{\beta}}\tilde{\theta}_j^{\dot{\beta}}), \ g\in \mathcal{G},
\end{equation}
where $V$, $\tilde{V}$ are invertible matrix functions of $g$.  Now restrict to a super null line $L$ tangent to a null vector $v^{\alpha\dot{\alpha}}=\lambda^{\alpha}\tilde{\lambda}^{\dot{\alpha}}$.  
Along $L$, the supersymmetry relations \eqref{eq:symintegrability} are preserved if and only if
\begin{equation}
v^{\beta\dot{\beta}}V^\alpha_{\beta}\tilde{V}^{\dot{\alpha}}_{\dot{\beta}}=((f\lrcorner L_0)_*v)^{\alpha\dot{\alpha}},
\end{equation}
where $L_0$ is the bosonic projection of $L$.    The existence of holomorphically varying matrix functions $V$ and $\tilde{V}$ satisfying this condition is the extra modification necessary to extend a causal morphism $f$ to $\hat{f}$, and will be assumed.

The above construction yields a symmetry of the $N=3$ SYM field equations, but we must also show that solutions of the YM field equations, with scalar and spinor fields set to zero, are preserved by these extended causal morphisms.  To do so, we will use two results proved by Harnad et. al.  \cite{harnad1989supersymmetric}.  In that paper, theorem 3.3 characterizes the form of the superconnection induced from a solution of the YM field equations when embedded as a gauge fixed N=3 SYM connection, which is 
\begin{align}\label{eq:nextcon}
\omega_{i\alpha}=\tilde{\theta}_i^{\dot{\alpha}}h_{\alpha\dot{\alpha}}(x^{\beta\dot{\beta}},\tau^{\beta\dot{\beta}}), \ \
\tilde{\omega}^i_{\alpha}=\theta^{i\alpha}\tilde{h}_{\alpha\dot{\alpha}}(x^{\beta\dot{\beta}},\tau^{\beta\dot{\beta}}), 
\end{align}
where $\tau^{\beta\dot{\beta}}=\sum_i \theta^{i\beta}\tilde{\theta}^{\dot{\beta}}_i$, $A_{\alpha\dot{\alpha}}=A_{\alpha\dot{\alpha}}(x^{\beta\dot{\beta}},\tau^{\beta\dot{\beta}})$, and the gauge condition is $\theta^{i\alpha}\omega_{i\alpha}+\tilde{\theta}^i_{\dot{\alpha}}\tilde{\omega}_i^{\dot{\alpha}}=0$.  Conversely, corollary 4.3 shows that a gauge-fixed connection that is integrable on super null lines and takes the above form corresponds to an $N=3$ extended solution of the YM field equations.  

Based on these considerations, showing that the extended causal morphisms preserve the form of equation \eqref{eq:nextcon} and the gauge condition implies that they preserve solutions of the YM field equations. To show this, restrict to a super null line $L$ and use \eqref{eq:sympullback} to compute the pullback connection of \eqref{eq:nextcon}, which gives
\begin{align}\label{eq:ompullback}
\begin{split}
&\lambda^\alpha \hat{f}^*\omega_{i\alpha}(\spoint)=\tilde{\theta}_i^{\dot{\beta}}\lambda^\alpha \left(V^\beta_{\alpha}h_{\beta\dot{\beta}}(x',\tau')\right), \\
 &\tilde{\lambda}^{\dot{\alpha}} \hat{f}^*\tilde{\omega}^i_{\dot{\alpha}}(\spoint) =\theta^{i\beta}\tilde{\lambda}^{\dot{\alpha}} \left(\tilde{V}^{\dot{\beta}}_{ \dot{\alpha}}\tilde{h}_{\beta\dot{\beta}}(x',\tau')\right),
\end{split}
\end{align}
where $x'$, $\tau'$ are evaluated at $\hat{f}\lrcorner L(\spoint)$ and $z\in \mbb{C}^{4\vert 4N}$.  Now consider two points $\spoint_1,\spoint_2\in \mbb{C}^{4\vert 4N}$ with $x_1=x_2$ and $\tau_1=\tau_2$, lying on two parallel lines.  Under $\hat{f}\lrcorner L$, $\tau$ transforms as $\tau^{\alpha\dot{\alpha}}\rightarrow [V^{-1}]^\alpha_\beta [\tilde{V}^{-1}]^{\dot{\alpha}}_{\dot{\beta}}\tau^{\beta\dot{\beta}}$, so the quantities in parentheses are the same for these two points and parallel lines, but the line can be varied, so this is true for any $(\lambda, \tilde{\lambda})$.  By linearity, this implies that the connection has the form of equation \eqref{eq:nextcon}, as desired. Furthermore, the gauge condition can be written $\tau^{\alpha\dot{\alpha}}(h_{\alpha\dot{\alpha}}+\tilde{h}_{\alpha\dot{\alpha}})=0$.  For $\tau^{\alpha\dot{\alpha}}$ proportional to a null vector this condition is preserved, but this must be true for any null line, so again by linearity the gauge condition is preserved.  This implies that the pullback connection corresponds to an $N=3$ extended solution of the YM field equations.

\section{Discussion}

We have shown that self-dual, $N=3$ super causal and causal morphisms yield  symmetries of the ASDYM, $N=3$ SYM and YM field equations, respectively.  To further understand these symmetries, it will be necessary to classify their solutions and to investigate their action on concrete examples of YM fields.  Some partial results were found in the previous paper \cite{https://doi.org/10.48550/arxiv.2203.07952}, where examples of self-dual morphisms were constructed from holomorphic endomorphisms of twistor space.  A method was also developed to construct causal morphisms from these self-dual morphisms.  Although these constructions provide examples of solutions, it will be important to find a more complete classification.  In particular, it is likely that there are more general examples than those constructed from endomorphisms of the twistor space, or super ambitwistor space, which could be analogous to holomorphic functions that preserve the real line in two dimensions. This preliminary interpretation is based on the CR ambitwistor space used in \cite{Mason_2006}, but will require further investigation to make precise. It will also be important to understand how these maps are related to other well known hidden symmetries for these equations.

There are many additional avenues of further research.  Here the action of these maps was only considered for classical Yang-Mills fields, but the ultimate goal is to further understand the quantum theory.  Furthermore, it will be interesting to consider how gravitational fields transform under these maps.  In this vein, one could define a causal manifold with coordinate transformations that are morphisms of these types, in analogy to the definition of Riemann surfaces for holomorphic functions.  Due to the nonlocal nature of these maps,  the theory could lead to interesting new mathematics.

\bibliographystyle{JHEP}
\bibliography{References}

\end{document}